# High-Q Cavity Interface for Color Centers in Thin Film Diamond

Sophie W. Ding*[1], Michael Haas*[1], Xinghan Guo*[2], Kazuhiro Kuruma[1,3], Chang Jin[1], Zixi Li[2], David D. Awschalom[2,4], Nazar Delegan[2,4], F. Joseph Heremans[2,4], Alex High[2,4], Marko Loncar[1]

[1]*John A. Paulson School of Engineering and Applied Sciences, Harvard University, Cambridge, Massachusetts 02138, USA*
[2]*Pritzker School of Molecular Engineering, University of Chicago, Chicago, IL 60637, USA*
[3]*Research Center for Advanced Science and Technology, The University of Tokyo, 4-6-1 Komaba, Meguro-ku, Tokyo 153-8505, Japan*
[4]*Center for Molecular Engineering and Materials Science Division, Argonne National Laboratory, Lemont, IL 60439, USA*
*wding@g.harvard.edu; loncar@seas.harvard.edu; ahigh@uchicago.edu*

## Abstract

Quantum information technology offers the potential to realize unprecedented computational resources via secure channels capable of distributing entanglement between quantum computers. Diamond, as a host to atom-like defects with optically-accessible spin qubits, is a leading platform to realize quantum memory nodes needed to extend the reach of quantum links. Photonic crystal (PhC) cavities enhance light-matter interaction and are essential ingredients of an efficient interface between spins and photons that are used to store and communicate quantum information respectively. Despite great effort, however, the realization of visible PhC cavities with high quality factor (Q) and design flexibility is challenging in diamond. Here, we demonstrate one- and two-dimensional PhC cavities fabricated in recently developed thin-film diamonds, featuring Q-factors of $1.8 \times 10^5$ and $1.6 \times 10^5$, respectively, the highest Qs for visible PhC cavities realized in any material. Importantly, our fabrication process is simple and high-yield, based on conventional planar fabrication techniques, in contrast to previous approaches that rely on complex undercut methods. We also demonstrate fiber-coupled 1D PhC cavities with high photon extraction efficiency, and optical coupling between a single SiV center and such a cavity at 4K achieving a Purcell factor of 13. The demonstrated diamond thin-film photonic platform will improve the performance and scalability of quantum nodes and expand the range of quantum technologies.

## Introduction

Diamond, as a host to atom-like defects with optically accessible long-lived spin qubits, has emerged as a compelling platform for applications in quantum sensing[1–4] and communication[5–8]. Among diamond color centers, nitrogen vacancies (NV)[1–5], silicon vacancies (SiV)[6,7,9–11], and tin vacancies (SnV)[8,12–14] are particularly promising, and have enabled many state-of-the-art demonstrations of quantum communication. Fundamentally, practical quantum communication requires fast and low-loss transfer of quantum information between spins (stationary qubit/quantum memory) and photons (flying qubit), which translates to figures of merit like bandwidth and fidelity. Therefore, achieving efficient spin-photon interfaces is crucial for a wide range of applications. In diamond, these interfaces have been realized by embedding the emitters within optical structures, including microcavities[15], microrings/microdisks[16], waveguides[8,11,17], and nanophotonic cavities[6,9,12,13,18,19], which have been enabled by steady progress in diamond fabrication techniques. Among all structures, photonic crystal (PhC) cavities are one of the most efficient spin-photon interfaces because they support optical modes with high quality factors (Q) and small mode volumes (V), which greatly enhances light-matter interactions and allow for efficient control and readout of the emitter spin state. As a result, PhC cavities have been utilized in a diverse range of qubit platforms, including quantum dots[20], defects in Si or SiC[21–23], and rare-earth ions in host materials [24–26].

Several methods have been developed to fabricate diamond nanophotonic structures from bulk diamond substrates, including the focused-ion-beam (FIB) milling[19], Faraday-cage angled etching[27], and approaches based on reactive-ion-beam (RIE) angled etching[6,9] (Figure 1a) or quasi-isotropic etching[12,13,18] (Figure 1b). Both angled and quasi-isotropic etching are the state-of-the-art methods and most popular as a result. However, despite great efforts, these methods typically result in visible PhC cavities with Q factors up to the low ~$10^4$ range, much lower than the simulated values >$10^6$, likely limited by fabrication imperfections such as roughness of etched surfaces induced by complex undercutting processes[28]. In parallel, diamond thin films generated by substrate thinning and ion-slicing followed by regrowth are explored as the tried-and true method, due to their potential to simplify and improve fabrication by eliminating undercutting. However, the PhC cavity Qs fabricated using these approaches have been limited to ~$10^3$, due to thickness variation and imperfect diamond crystal quality[29–31]. They are usually incompatible with quantum networking demonstrations for practicle reasons, so the bulk machining methods are still the work horse in the quantum diamond photonics community today. Therefore, a new thin-film platform and fabrication pathway that improve fabrication capability and maintain high diamond quality are needed to enhance the performance of photonic components in diamond.

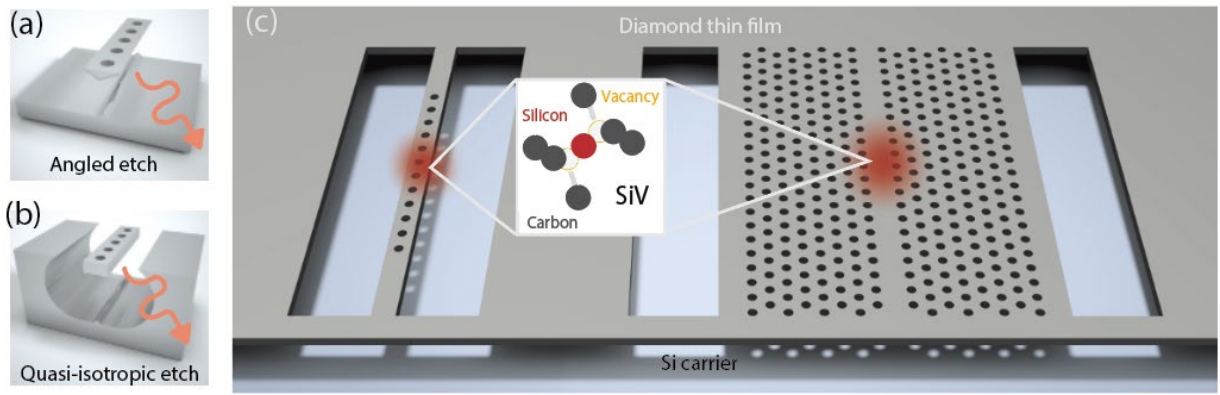

**Fig 1. State-of-art high-Q suspended PhC cavities fabrication methods.** (a) Angled etch of a 1D PhC cavity. The structure has a characteristic triangular cross section due to the nature of the fabrication process. Due to the shallow angle of the ion beam which would graze the top edge of nearby structures, dense patterns are prohibited. (b) Quasi-isotropic etch of a 1D PhC cavity. The suspension is achieved by an isotropic $O_2$ reactive-ion etch process. The disadvantage of this method is long undercutting time that scales with beam width. In both (a) and (b), the structure is carved out of a bulk diamond, and usually has visible artifacts/roughness on the bottom of the nanobeam due to the undercutting. (c) This work: thin film fabrication of 1D and 2D PhC cavities. This approach relies on top-down etching and undercutting of the handle substrate, so it simplifies the fabrication and avoids the roughness of the bottom surface. As a result, this approach leads to exceedingly higher Qs and also allows for more versatile photonic circuits to be realized.

In this paper, we demonstrate a new thin film diamond photonic platform and realize optical cavities in the visible wavelength range featuring record-high Q factors coupled to stable SiV centers. In our approach, we use a high-quality and homogenous thin film diamond bonded to a silicon oxide/silicon handle wafer. The film is created through ion implantation in bulk diamond, regrowth, electrochemical etching, and transfer printing[32]. We design and fabricate 1D and 2D PhC diamond cavities operating at 737 nm wavelength range and measure a Q-factor up to $1.8 \times 10^5$, a record for visible PhC cavities fabricated in any material (Table 1). We also fabricate 1D PhCs critically coupled to a feeding waveguide and measure loaded Q of $8.4 \times 10^4$ (intrinsic $Q \sim 1.8 \times 10^5$) and a waveguide-cavity coupling efficiency of ~ 65%. We show that this fabrication method exhibits high-yield and uniformity: 93% (53 out of 57) of the cavities feature high-Q modes with resonances matched closely to the designed resonances. Finally, we demonstrate coupling of implanted SiVs to fabricated diamond cavities and observe three-fold reduction of their radiative lifetime, achieving a Purcell factor of 13. We expect the exceptional cavity performance, high-yield fabrication process, and excellent SiV properties of this new platform to further advance the field of quantum photonics, as efficient spin-photon interfaces for color centers in diamond and beyond.

# Result

**Device fabrication**

In our approach, we begin by generating diamond thin films by ion slicing and overgrowth and then use a transfer printing process to directly bond the thin films to a SiO2/Si substrate[32]. The size and thickness of the diamond film are ~200x200 um and 160 nm, respectively. The film has a surface roughness < 0.3nm and thickness variation ~1 nm, both essential for minimizing optical scattering losses and achieving uniform, high-Q cavities. The SiVs are generated through implantation across the membrane before the transfer, resulting in randomly distributed SiVs. The fabrication process used to realize PhCs in diamond thin films is summarized in Fig. 2 (a). Scanning electron microscope (SEM) images of fabricated 1D and 2D PhC cavities are shown in Fig. 2 (b) and (c). The 1D cavities (lattice constant $a_{1D}$ = 184~226 nm and hole radius $r_{1d}$ = 65 nm) are formed by introducing a quadratic hole shift near the waveguide center[12], while the 2D cavities ($a_{2D}$ = 236~269 nm and $r_{2d}$ = 65 nm) are formed by shifting the center holes outwards in the PhC line-defect waveguide[29]. The details of 1D and 2D cavity designs and resulting field profiles can be found in Methods. The typical simulated Q and V for our 1D (2D) cavities are $\sim 1.0 \times 10^6$ ($\sim 7.6 \times 10^5$) and $\sim 0.5$ ($\sim 2.9$) $(\lambda/n)^3$.

For practical applications it is important to interface the cavity with a waveguide and eventually a fiber to efficiently transfer quantum information between local spins and propagating photons. Therefore, we also fabricate 1D PhC coupled to a feeding waveguide Fig. 2(e) by reducing the number of holes in the photonic crystal mirror on one side of the waveguide (see Methods). This results in the light preferentially coupling out of the cavity from the feeding waveguide and leading to a simulated cavity Q of $1.9 \times 10^5$ (for $a$ = 255 nm). The waveguide is tapered at the end (< 60 nm tip width) to allow efficient coupling to a tapered fiber [33,34].

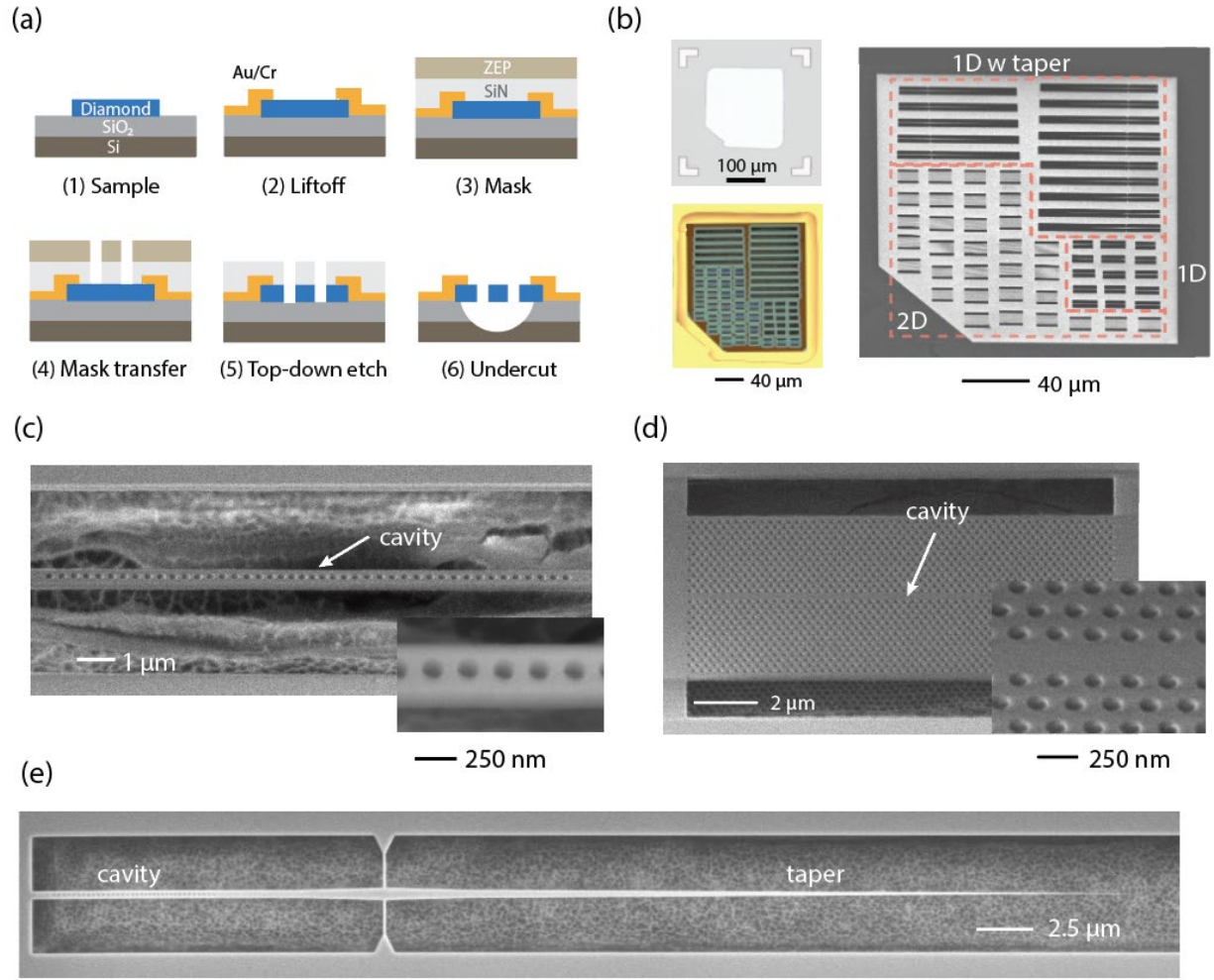

**Fig. 2: Fabrication of high-Q PhC cavities on a thin-film diamond membrane.** (a) Fabrication flow of the devices based on a thin-flim diamond platform. : (1) The diamond with implanted SiV is bonded to the SiO2/Si substrate; (2) A liftoff of Cr/Au metal stack is performed to define a "frame" around the film in order to secure it on the carrier chip; (3) a silicon nitride (SiN) hard mask is deposited using plasma-enhanced chemical vapor deposition (PECVD) with electron beam (EB) resist ZEP520A on top; (4) The cavity pattern is written using EB lithography, and is transferred onto the SiN hard mask using plasma reactive ion etching (RIE) in $SF_6$ and $H_2$ chemistry; (5)After removing the EB resist, the pattern is etched into diamond using RIE in $O_2$ chemistry; (6) Finally, hydrofluoric acid (HF) is used to remove the SiN mask and the 1-um oxide sacrificial layer underneath. To introduce additional separation between diamond film and underlying substrate, the handle wafer is exposed to $XeF_2$ to etch a deep trench (> 1um) under the membrane to avoid optical losses via coupling between the diamond membrane and silicon substrate . (b) Left top and bottom panels are microscope images of the sample before and after fabrication. Right panel is the SEM image of the fabricated sample. (c) The SEM image of the 1D PhC cavity. The inset shows the details of the holes. Both are taken at 45 degee angle. (d) The SEM image of the 2D PhC cavity. The inset shows the details of the holes. Both are taken at 45 degee angle. (e) The SEM image of the 1D PhC cavity with the taper. The geometry is formed by removing 9 holes on the one side from the symmetric 1D cavity design to allow for more preferential coupling to the tapered-waveguide side of the structure.

## Device characterization: 1D and 2D photonic crystal

The fabricated devices are first characterized by photoluminescence (PL) measurements at room temperature (see Method). Figures 3 (a) and (b) show the PL spectra measured for 1D and 2D PhC cavities with different lattice constants, $a_{1D/2D}$. The broad peak at 737 nm is the emission of SiV centers. For 1D PhC cavities, we observe sharp peaks corresponding to the fundamental and high-order (2nd and 3rd) modes. A clear red-shift of their resonant wavelengths is observed as the values of $a$ increase. The measured wavelengths of fundamental modes are in good agreement with the simulations (see the upper panel of Fig. 3 (c)). 2D PhC cavities also exhibit peaks corresponding to the band edges and fundamental modes, and show shift of the cavity wavelengths as $a$ becomes larger, which is consistent with the simulation as well (see the lower panel of Fig. 3 (c)).

This fabrication method exhibits high uniformity (resonances close to simulation) and high-yield (high-Q modes) for both 1D and 2D PhC cavities: from Fig. 3 (c), we estimate the deviation of the measured wavelengths from the simulation to be within 2.9% and 2.5% for 1D and 2D cavities, respectively. We fit the cavity spectra observed in Figs. 3 (a) and (b) find that 100% (30 out of 30) of all 1D cavities and 85% (23 out of 27) of all 2D cavities characterized feature spectrometer-limited quality factors of $Q > \sim 2 \times 10^4$.

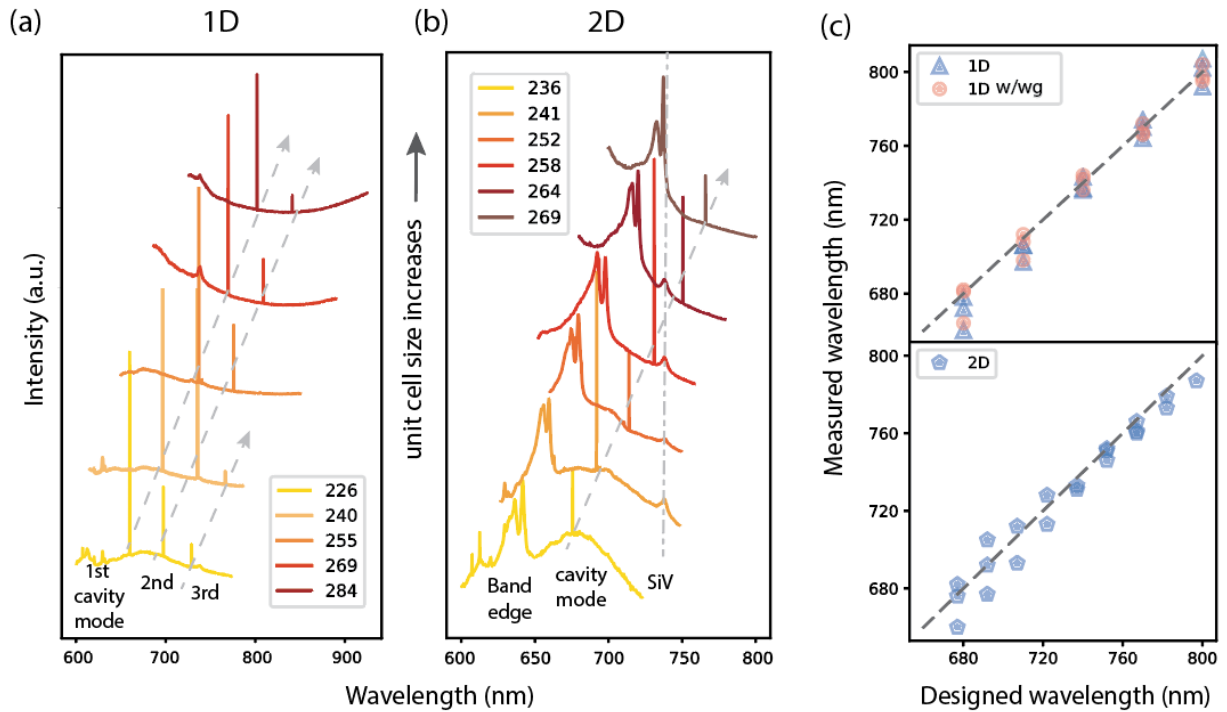

**Fig. 3: Lattice constant dependence of 1D and 2D PhC cavities:** Off-resonance excitation and spectrometer characterization of (a) 1D and (b) 2D phCs. The legend shows the unit cell length, in nm. The intensity is plotted against the wavelength, showing resonance features of the cavities and the SiV at 737 nm. The spectra are offset for clarity. The legend shows the corresponding design lattice constant $a$, in nm. The cavity modes redshift as the lattice constant increases, as expected. (c) The measured cavity resonances

(fundamental mode) versus the designed ones for 1D, 1D with taper, and 2D PhC cavities presented in this paper. We observe excellent agreement between the two.

To further resolve the high-Q cavity resonances, we perform cross-polarized measurements using a tunable CW laser[29], as schematically shown in Fig. 4 (a). We scan the laser across the cavity resonance and detect the scattered light from the sample by an avalanche photodiode (APD). Figures 4 (c) and (d) show the reflection spectra of the fundamental cavity modes measured for 1D and 2D cavities with the highest Q factors. The peaks are fitted to a Lorentizian function, and the Q factors obtained are $1.83 \times 10^5$ and $1.60 \times 10^5$, respectively. We note that the Q value for 1D PhC device is one order of magnitude higher than those previously reported in visible diamond PhC cavities, setting a new record for visible PhC cavities fabricated in any material (Table 1). Furthermore, our method is compatible with fabrication of state-of-the-art 2D PhC cavities in diamond that are difficult to make using bulk machining approaches or previous thin film diamond approaches: we demonstrate 2D PhC devices with Q factors that are 20~100 times higher than those shown before (Table 1). Lastly, most measured 1D and 2D cavities on the same membrane exhibit Q factors over $0.5 \times 10^5$ as shown in Fig. 4 (f). which further showcases the uniformity and yield of this platform.

Furthermore, our method is compatible with fabrication of state-of-the-art 2D PhC cavities in diamond that are difficult to make using bulk machining approaches or previous thin film diamond approaches: The highest Q factor achieved in the 2D PhC device is also we demonstrate 2D PhC devices with Q factors that are 20~100 times higher than previous values reported using quasi-isotropic etching and thin film based approach those shown before (Table 1), suggesting the usefulness of our platform for the realization of high experimental Q factors. Lastly, most measured 1D and 2D cavities on the same membrane exhibit Q factors over $5 \times 10^4$ as shown in Fig. 4 (f). which further showcases the uniformity and yield of this platform.

To characterize the 1D PhC cavities coupled to a waveguide, we use an optical fiber coupling system, as shown in Fig. 4 (b). The light from the tunable laser is sent to the tapered waveguide through the optical fiber and the reflected light is detected by an APD. We use a low laser power of ~pW to avoid distortion of the cavity spectrum due to the thermo-optic effect (see Methods). Fig. 4 (e) shows the cavity spectrum of waveguide-coupled 1D PhC cavity with the highest Q factor. The loaded Q factor is measured to be $8.4 \times 10^4$. The cavity dip shows full contrast (95.3±0.3 % contrast), indicating that our cavity is almost critically coupled. This allows us to estimate the intrinsic Q to be $Q_i = (1.8 \pm 0.4) \times 10^5$. The coupling efficiency between the cavity and the waveguide is measured to be $\eta_c \sim 65\%$ in our system (see Methods).

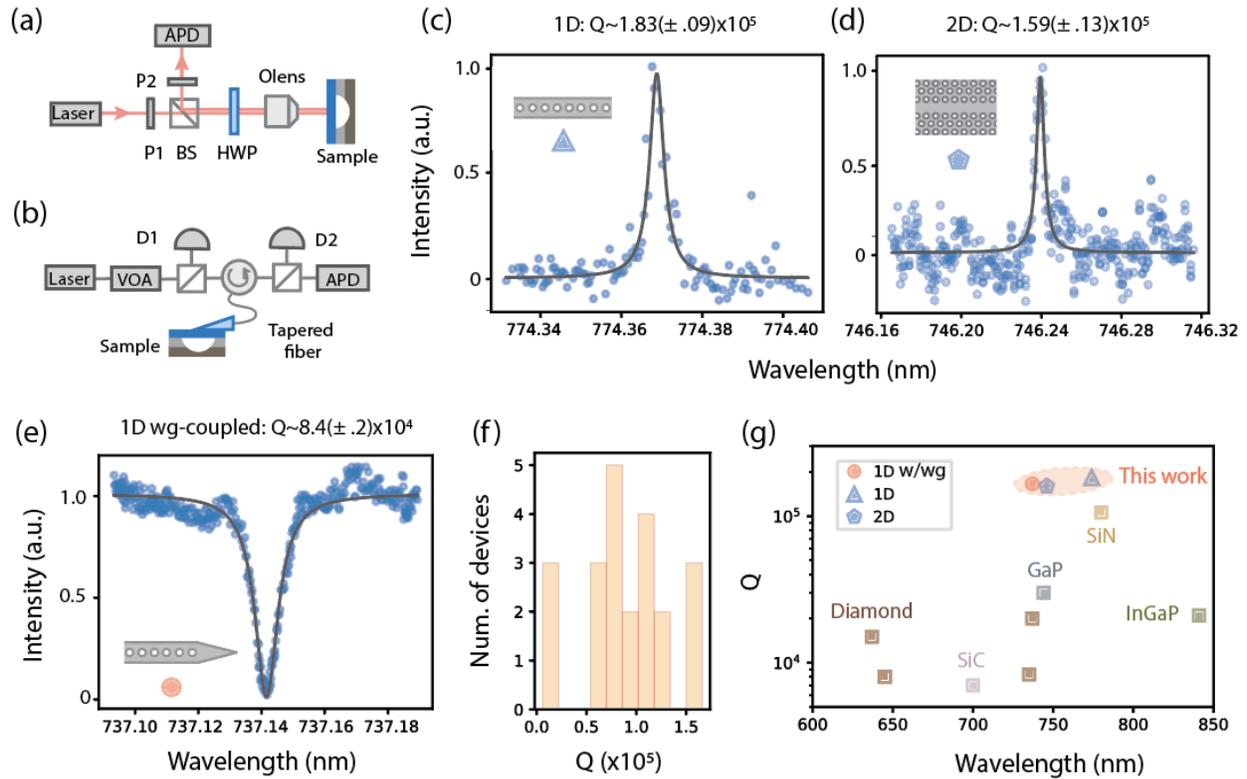

**Fig. 4: 1D and 2D PhC cavity on-resonance characterizations for measuring quality factors:** (a) The cross-polarization setup for on-resonance characterization of the cavity. Olens: objective lens; HWP: half wave plate; BS: beam splitter; P1, P2: polarizers. (b) The fiber-coupling setup for on-resonance characterization of the cavity. VOA: variable optical attenuator; D1, D2: detectors. Laser: visible (710~790 nm) scanning laser; APD: avalanche photodiode. (c)(d) On-resonance scan of the 1D and 2D cavity using the cross-polarization setup. (e) On-resonance scan of the 1D cavity coupled to the feeding waveguide. The cavity is critically coupled as the contrast is ~95% in reflection. (f) Histogram of all the PhC cavities measured and resolved in the 710~770 nm range. (g) PhC cavity measurement summary. The values from this work are plotted along with the highest literature values in their respective categories, as shown in Table 1.

## Characterization of SiV properties

Stable and bright emitters are required for cavity quantum electrodynamics (QED) experiments. To demonstrate the immediate compatibility of this platform for efficient photonic interface for emitters embedded in diamond, we characterize the SiVs in the fabricated devices at 4K and demonstrate optical coupling between an SiV and a high-Q 1D PhC cavity.

We optically characterize two single SiV centers embedded in the realized high-Q cavities by confocal PL measurements at 4K, as shown in Fig 5. (a). The spectrum of an SiV at 4K features four optical transitions labeled from A to D around 737 nm, as shown in Fig 5. (b). We first characterize the C transition of a single SiV at 737.09 nm. Figure 5 (d) shows the color map of its photoluminescence excitation (PLE) spectra measured over two hours (upper panel) and a single scan spectrum (lower panel). There is no significant spectral diffusion and the linewidth is fitted to be 605 MHz. The observed linewidth is slightly larger than the typical value of ion-implanted SiV⁻ centers[35], possibly due to the laser broadening or/and membrane/cavity fabrication[36]. We then perform the second-order correlation ($g^2(\tau)$) measurements using a Hanbury Brown-Twiss setup (see Methods). The measured intensity correlation histogram (Fig. 5 (c)) exhibits a clear antibunching with a value of the zero time delay $g^2(0)$ of 0.31±0.12, confirming the single-photon nature of the investigated SiV. The non-zero value of $g^2(0)$ could be due to dark counts of the APD, drift of the sample position, and diffusion of the ZPL line over a long time.

We finally investigate the optical coupling between a single SiV center and a 1D cavity in our thin film diamond platform. We choose a cavity device with measured resonance at 737.5 nm and experimental Q of $1.2 \times 10^5$. The measured PL spectra of the cavity device features four sharp peaks, corresponding to A ~ D transitions of an SiV, and the cavity resonance which is gradually tuned to the D line using a gas condensation approach[12] (Fig. 5 (e)). The SiV experiences low strain, indicated by the ~55 GHz splitting between C and D line[37]. We observe small shifts of the D line during the tuning, which could indicate changes in strain environment. It is noted that the linewidth of the cavity spectra shown is limited by our spectrometer resolution of ~10 GHz. Under the resonance condition (spectral detuning between the D line and cavity of $\Delta \sim 0$ nm, red spectrum in Fig. 5(e)), we observe a ~20-fold intensity enhancement of the D line emission compared to that under far-detuned conditions ($\Delta \sim 0.4$ nm, blue spectrum in Fig. 5(e)), suggesting that the spontaneous emission rates of the SiV center are enhanced by Purcell effect.

We also conduct time-resolved PL measurements on the same SiV center (see Method). Figure 5 (f) shows the PL curves measured under the resonance (red) and far-detuned condition (blue). By fitting the curves with a single exponential function, we obtain the lifetime of the SiV on resonance ($\tau_{on}$) to be 0.47 ± 0.006 ns, which is reduced approximately by a factor of 3 from the off-resonance value ($\tau_{off}$) of 1.3 ± 0.01 ns. The Purcell factor of the investigated zero-phonon line (ZPL) $F_{ZPL}$ is estimated to be 13 (see Methods). This value is much smaller than the theoretical Purcell factor (see Methods), likely due to the large displacement of SiV position with respect to the cavity field maximum as a result of random bulk implantation. To further improve the Purcell factor, the SiVs can be implanted at the cavity region using the well-established masked implantation technique[6,7]. We estimate that for SiV mask-implanted at the cavity regions, the cooperativity $C$, an important figure-of-merit for evaluating the emitter-cavity coupled system, could reach > 960 (undercoupled, $Q \sim 1.8 \times 10^5$) or > 440 (critically coupled, $Q \sim 8.4 \times 10^4$) with experimental values (see Methods), which are 9 or 4 times higher than the highest value reported in a previous work[6]. The realization of such a large $C$ as a result of the high-Q cavity is highly advantageous for high fidelity of control of the spin/nuclear states and networking applications.

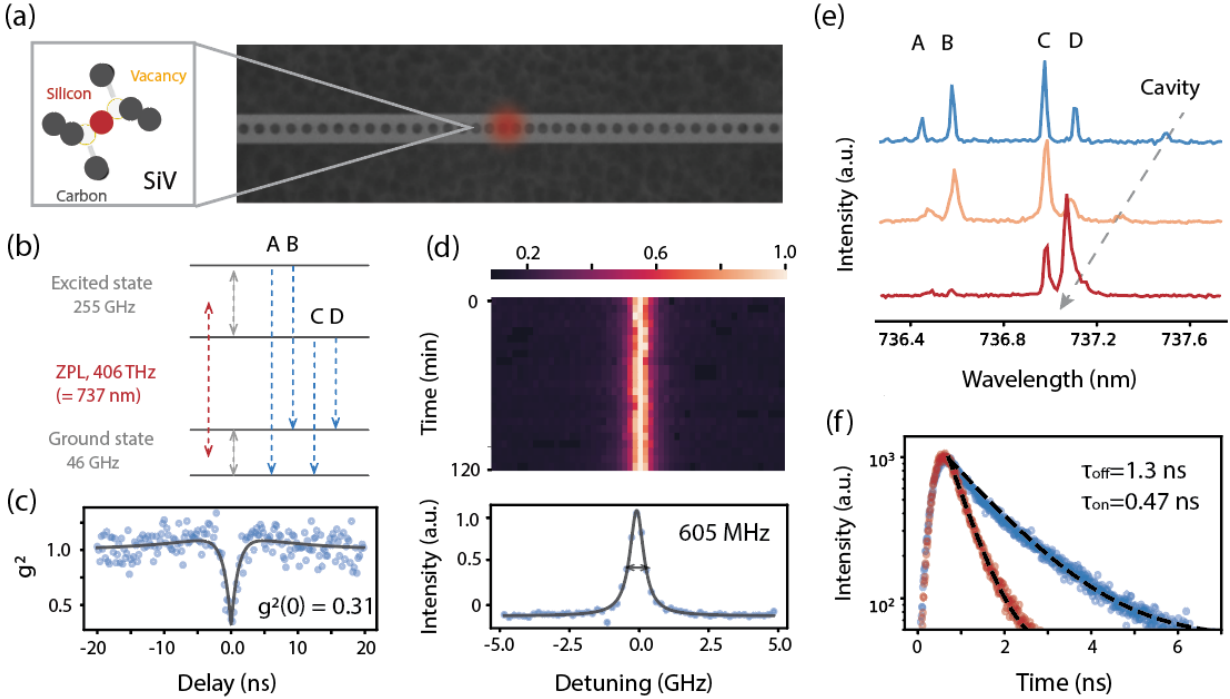

**Fig. 5: SiV characterization and its optical coupling to the PhC cavity at 4 K:** (a) An illustration of SiV placed close to the center of PhC cavity (SEM image). SiV is an interstitial defect, consisting of Si atoms positioned between two sites with missing carbon atoms in diamond lattice. (b) A simplified level diagram of SiV, featuring A~ D, 4 optical transitions around 737 nm. (c) Second-order autocorrelation measurement of the C line under on-resonant excitation. The $g^2(0) < 0.5$ confirms the single-photon nature of the emitter. (d) Top panel: the stability of the C line of an SiV over two hours. The plot shows normalized intensity over time. The drift is much less than a linewidth; bottom panel: a scan of the C line at 120 min. (e) The photoluminescence (PL) spectra of the SiV as the cavity is tuned closer to the C and D lines. Each spectrum is normalized to the highest peak. The D line is getting much brighter as the cavity is tuned into resonance. (f) The measured lifetime of the SiV on and off resonance (bulk) with the cavity, which shows a factor of 3 reduction. The two lifetimes correspond to the initial and final PL spectra in (e) of the same colors.

## Summary and conclusions

We have demonstrated high-$Q$ 1D and 2D PhC cavities using the thin film diamond approach. We achieved Q factors up to $1.8 \times 10^5$, which is a record high in visible-wavelength PhC cavities in any materials. We also fabricated high-Q 2D PhC cavities that were previously challenging using conventional diamond bulk machining approaches, and also achieved record-high Q values up to $1.6 \times 10^5$. Finally, of interest for practical applications and demonstrations of efficient spin-photon interface, we demonstrated waveguide-coupled 1D PhD cavities featuring intrinsic (loaded) Q up to $1.8 \times 10^5$ ($8.3 \times 10^4$) and coupling efficiency of 65%. The significant improvement of the experimental Q factors in this work can be attributed to the use of a high-optical-quality thin film diamond membrane with a smooth surface roughness < 0.3nm and small thickness variation ~1 nm [32]. Still, Qs are one order of magnitude lower than theoretical predictions, indicating that there is room for improvement. The difference could be due to surface absorption[38] and/or optical scatterings due to fabrication imperfections such as lithography error of the air holes in position or radius, surface roughness, sidewall roughness/tilt of air holes[39].

Lastly, we demonstrated optical coupling between the realized high $Q$ cavity and a single SiV, with measured Purcell factor of 13, and therefore the immediate compatibility of this platform with color-center cavity QED. Using masked implantation, better overlap between SiV and optical mode can be achieved, resulting in cooperativities > 440. The lower loss and design flexibility in this platform can fundamentally enhance color-center-based technologies, by allowing for higher single photon rates, higher gate fidelities, more integrated functionalities, etc. In combination with the robust, high uniformity, and high-yield fabrication process, our thin-film diamond platform will hopefully unlock new opportunities for color-center applications for quantum information.

Our platform and fabrication approach based on a thin-film diamond can be applied to a variety of other micro/nanostructures that are important in diamond integrated photonic applications such as nonlinear photonics[40,41] and diamond phononics[42,43]. Additionally, the flexibility of direct bonding makes this approach easily applicable to other substrates besides Si/SiO$_2$[32], which enables the heterogeneous integration of the diamond platform containing color centers onto existing and emerging integrated photonic circuits for quantum networks, including thin-film lithium niobate[11], aluminum nitride[10], and CMOS-compatible platforms[44].

| Cavity type, material | Wavelength (nm) | Q | V ($\lambda$/n)$^3$ | Method | Reference |
|---|---|---|---|---|---|
| **1D, diamond** | **737** | **8.3x10$^4$/1.8x10$^5$** | **0.5** | **Thin film** | **This work** |
| 1D, diamond | 637 | 1.4x10$^4$ | ~1 | Quasi-isotropic etching | Mouradian[18] |
| 1D, diamond | 737 | 2.0x10$^4$ | 0.5 | Angle etching | Bhaskar[6] |
| 1D, diamond | 660 | 2.4x10$^4$ | 0.5 | Photoelectrochemical etching | Lee[31] |
| 1D, diamond | 1529 (telecom) | 1.8x10$^5$/2.7x10$^5$ | 0.57 | Angle etching | Burek[42] |
| 1D, SiN | 780 | 1.1x10$^5$ | 0.4 | Thin film | Samutpraphoot[45] |
| 1D, AlN | 403 | 6.9x10$^3$ | 1.6 | Thin film | Sergent[46] |
| 1D, 4H-SiC | 700 | 7x10$^3$ | 0.5 | Photoelectrochemical etching | Bracher[47] |
| 1D, GaP | 744 | 3.0x10$^4$ | ~1 | Monolithic | Chakravarthi[48] |
| 1D, InGaP | 841 | 2.1x10$^4$ | 0.64 | Monolithic | Saber[49] |
| **2D, diamond** | **746** | **1.6x10$^5$** | **2.18** | **Thin film** | **This work** |
| 2D, diamond | 645 | 8x10³ | 0.35 | Fib | Jung[19] |
| 2D, diamond | 1470 (telecom) | 1.8x10³ | 2.15 | Thin film | Kuruma[29] |

**Table 1**: Visible and telecom wavelength suspended diamond PhCs, and visible wavelength suspended PhCs in other low loss visible photonic materials.

## Method

### Device design

The 1D and 2D PhCs in this paper are simulated using a 3D finite-difference time-domain (FDTD) method. They share the same geometry as in a previous work[12]. Fig. S1 (a) and (b) show the mode profile and the geometry of the cavity region of the 1D PhC cavity. Fig. S1 (a) inset shows the mode profile of the 1D PhC device coupled to a feeding waveguide. The free-standing diamond nanobeam is designed to be 370 nm wide (w), and 160 nm thick (d). For the mirror cells, we scan the lattice constant $a$ from 226 nm to 284 nm, to cover a wide range of target wavelength from 680 nm to 800 nm of the cavity. The airhole radius (r) is 65 nm. To form the cavity, we taper the lattice constant at the center quadratically, with $a_1 = 0.84a$, $a_2 = 0.844a$, $a_3 = 0.858a$, $a_4 = 0.88a$, $a_5 = 0.911a$, $a_6 = 0.951a$. Fig. 1 (b) shows the electric field profile ($E_y$) of the fundamental mode of the cavity with $a = 269$ nm. The simulated Q is 1.3 x 10⁶, with a mode volume of $V \sim 0.5(\lambda/n)^3$. The waveguide-coupled devices are designed by removing 9 holes from one side of the 1D PhC device. The holes that form the cavities have the same dimensions. When $a$ is set to be 255 nm for the devices with tapered waveguides, the simulated Q is 1.9 x 10⁵, with a mode volume of $V \sim 0.5(\lambda/n)^3$. The coupling efficiency between the cavity and diamond waveguide $\eta_c$ is simulated to be ~87% by FDTD simulations.

Fig. S1 (c) and (d) show the mode profile and the geometry of the cavity region of the 2D photonic crystal. The cavity is created by the local width modulation of a line defect. The lattice is triangular, where the lattice constant (a) ranges from 236 nm to 269 nm to target wavelength from 677 to 767 nm, and the hole radius (r) is also 65 nm. By keeping the air hole size constant across a given chip, we can minimize systematic distortion in the hole sizes in fabrication. The cavity is formed by removing shifting the central holes "outwards" towards the mirror region in a linear fashion, where $b_1 = 10.1$ nm, $b_2 = 0.75b_1$, $b_3 = 0.5b_1$, $b_4 = 0.25b_1$. Fig. 1 (d) shows the electric field profile ($E_y$) of the fundamental mode of the cavity with $a = 252$ nm. The simulated Q is 6.3 x 10⁶, with a mode volume of $V = 2.9(\lambda/n)^3$

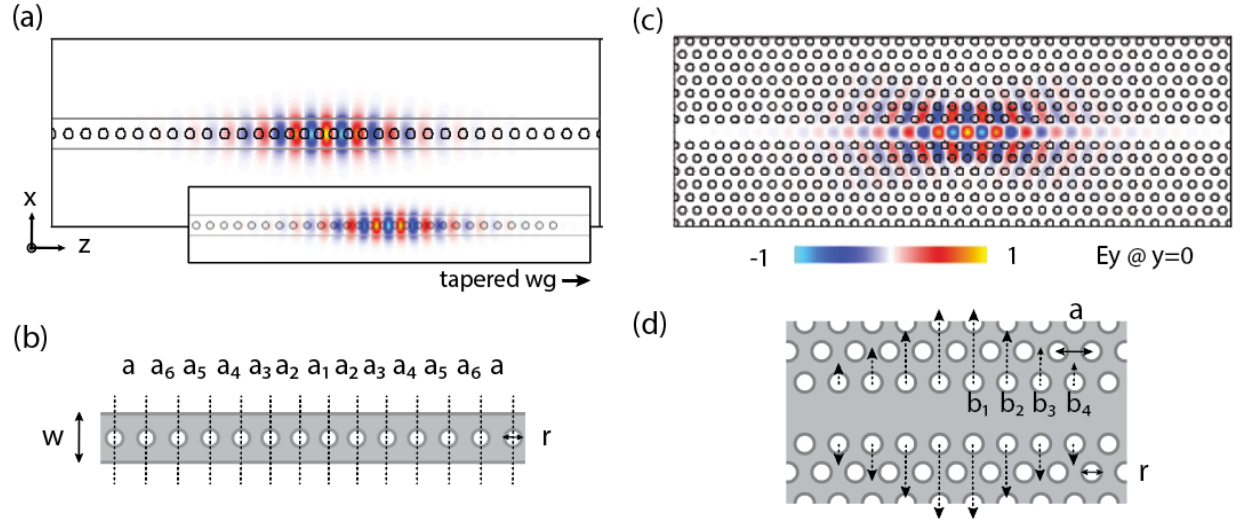

**Extended data Fig. 1** (a) 1D phC mode profile ($E_y$). The field is normalized. (b) 1D phC design. The cavity is formed by tapering the lattice constant quadratically, $a_1$ to $a_6$. (c) 2D phC mode profile ($E_y$). (d) 2D phC design. The cavity is formed by shifting the holes, by $b_1$ to $b_4$.

## Measurement setup
### Confocal photoluminescence measurements
In Fig.3, both spectra for 1D and 2D devices are measured using a commercial spectrometer system with free-space off-resonance excitation and collection. A green diode laser (523 nm) is used to scatter off of the cavities via an objective (x100), and the scatter light is sent through a 1800 gr/mm grating and collected by a Si CCD camera cooled to 4K. The spectrometer used is SpectraPro HRS-750, the supercontinuum laser is SuperK EXTREME.

### Cross-polarized measurements
In Fig. 4, the visible scanning laser is M Squared, which can stabilize in 710-790 nm, and the APD is SPCM-AGRH-14-FC from Perkin Elmer. For these measurements, the input polarizer is aligned to the cavity polarization (TE), the half-wave plate is set to 23.5° so that the laser polarization is 45° relative to the cavity polarization, and the output polarization is set to be perpendicular to the cavity polarization. The working principle behind the measurement is that the cavity acts as a polarization filter when the incident light is resonant with the cavity. As such, off-resonant light is fully blocked by the cross polarization between P1 and P2, but light resonant with the cavity has a non-zero polarization component in the direction of P2, resulting in a Lorentzian peak in reflection corresponding to the cavity resonance.

### Fiber-coupling measurements
The waveguide-coupled 1D PhC cavities are measured via fiber-coupled reflection measurements (shown in Figure 4 e). These measurements are performed by inputting light from a tunable laser (M Squared) into the diamond waveguide via a tapered optical fiber (S630 HP)[33] and measuring the reflected light from the optical cavity. We can estimate the degree of over/under coupling via the depth of the cavity reflection dip

$R_0$ via the formula $\kappa_i = \frac{\kappa(1 \pm \sqrt{R_0})}{2}$ where the plus (minus) refers to the cavity being over (under) coupled. For the cavities, where $R_0$ is close 0, this is a signature of the cavities being near-critically-coupled ($\kappa_e \sim \kappa_i$). For the fitted 95.4% contrast, the $Q_i$ and $Q_e$ are estimated to be $1.4 \times 10^5$ and $2.2 \times 10^5$, interchangeably.

In order to calculate coupling efficiency, it is necessary to decouple setup-related losses from losses related to the fiber-device coupling interface. We can express this mathematically by writing the total loss from D1 to D2 ($\eta_{tot}$) as a product of these losses, i.e. $\eta_{tot} = \eta_s \eta_c^2$. To find $\eta_s$, we splice a retroreflector into the setup at port 2, such that the loss from D1 to D2 can be solely attributed to setup losses $\eta_s$. Once this is calibrated, we can attribute any excess losses in the device measurement as fiber-device coupling losses.

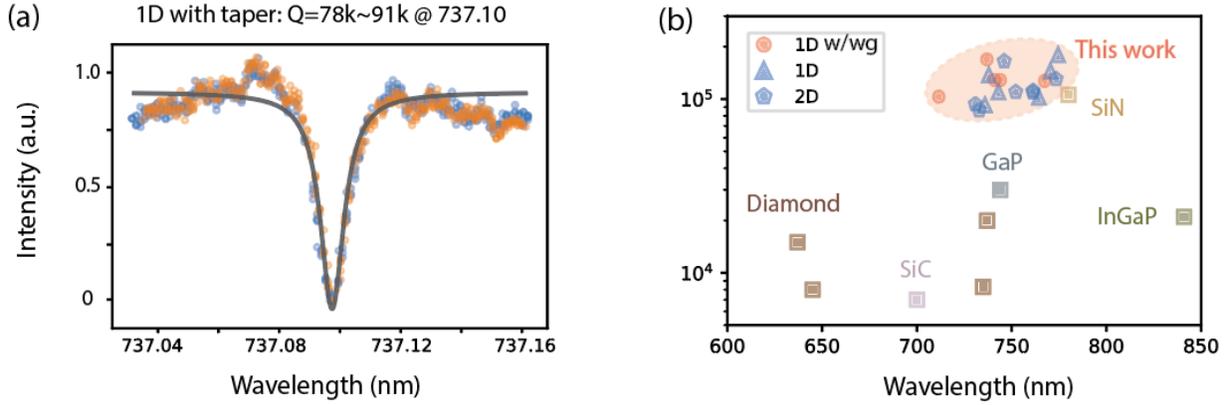

**Extended data Fig. 2 Q measurement summary.** (a) The forward and backward scan of the waveguide-coupled 1D PhC shown in the main text. (b) A summary of the Qs obtained from on-resonance laser scans for all devices on the sample, compared with the highest literature Q factors in diamond 1D and 2D PhC.

To confirm that the peaks are not broadened or narrowed because of thermo-optics caused by the laser sent into the cavity, we scanned both forward and backward with the same scanning speed and laser power. The fitted Q is between $7.8 \sim 9.1 \times 10^4$, as shown in Fig. S3 (a).

We can resolve the devices with resonances reachable by the scanning laser, in the range of 710-790 nm. They are plotted in Fig. S2 (b) against literature values presented in Table 1, which is an extended version of Fig. 4. (g) in the main.

**Calculating Purcell Factor and Cooperativity**

The Purcell factor of the investigated zero-phonon line (ZPL) $F_{ZPL}$, is estimated using the following equation[9]: $F_{ZPL} = (\tau_{off}/\tau_{on} - 1)/\varepsilon_{ZPL}$ [9], where $\varepsilon_{ZPL}$ is defined by the fraction of the total emission into the ZPL visible at 4K for an SiV, which is estimated by a product of the Debye–Waller factor of 70%[50] and the branching ratio of 19.3% into D line at 4K[9]. For this estimation, we also assume that the lifetime of a SiV measured in unpatterned area $\tau_{bulk}$ (~1.2 ns) is equal to the off-resonance value[9].

The theoretical Purcell factor is obtained from the following equation: $F \equiv \frac{3}{4\pi^2} \frac{\lambda^3}{n^3} \frac{Q}{V}$. We use the measured $Q \sim 1.2 \times 10^5$ and simulated $V \sim 0.5(\lambda/n)^3$, and estimated $F$ to be $1.8 \times 10^4$ in the most ideal case.

The cooperativity is calculated using $C = 4g^2/(\kappa\gamma)$, where $g$ is the coupling rate between the SiV and the cavity, $\kappa$ and $\gamma$ are the decay rate of the cavity and the emitter, respectively. If the SiV is placed at the cavity field maximum, the estimated $g$ is 15.2 GHz, and experimentally, it has been observed to be ~8 GHz[6]. To be realistic, we use $g \sim 8$ GHz, the natural SiV linewidth ($\gamma \sim 0.12$ GHz)[6], and the measured highest $Q$ of $1.8 \times 10^5$ ($\kappa \sim 2.2$ GHz) or Q of $8.4 \times 10^4$ ($\kappa \sim 4.8$ GHz) in the critically-coupled case. The obtained values are $C \sim 970$ or $C \sim 440$. If we use the theoretical $g \sim 15.2$ GHz, the C could be as large as 3500.

## Acknowledgement

The authors thank Dr. C. De-Eknamkul, Dr. B. Pingault, and Dr. R. Katsumi for helpful discussions. The research is funded through AWS (Partnership for Quantum Networking), NSF ERC (EEC-1941583), ONR (N00014-20-1-2425), AFOSR (FA9550-20-1-0105 and MURI on Quantum Phononics), ARO MURI (W911NF1810432). The membrane synthesis is funded through Q-NEXT, supported by the U.S. Department of Energy, Office of Science, National Quantum Information Science Research Centers. The membrane bonding work is supported by NSF award AM-2240399 and made use of the Pritzker Nanofabrication Facility (Soft and Hybrid Nanotechnology Experimental Resource, NSF ECCS-2025633) and the Materials Research Science and Engineering Center (NSF DMR-2011854) at the University of Chicago. Diamond growth related efforts were supported by the U.S. Department of Energy, Office of Basic Energy Sciences, Materials Science and Engineering Division (N.D.)